\documentclass[aps,floats,twocolumn,epsf,prb,showpacs]{revtex4-1}

\usepackage{graphicx}

\usepackage{color}
\usepackage{MnSymbol}
\usepackage{nicefrac}
\usepackage{braket}
\usepackage{array}
\usepackage{bm}

\begin{document}

\title{Nickel-Titanium double perovskite: \\ A three-dimensional spin-1 Heisenberg antiferromagnet}
\author{M. Karolak, M. Edelmann and G. Sangiovanni}
\affiliation{Institut f\"ur Theoretische Physik und Astrophysik, Universit\"at W\"urzburg, Am Hubland, D-97074 W\"urzburg, Germany}

\date{ \today }

\begin{abstract}
The double perovskite La$_2$NiTiO$_6$ is identified as a three-dimensional $S$=1 quantum magnet. 
By means of Density Functional Theory we demonstrate that this material is a high-spin $d$-electron system deep in the Heisenberg limit and establish that its paramagnetic Mott phase persists down to low temperatures (experimental N\'eel temperature $T_{\rm N}\sim 25$K) not because of frustration effects but rather for strong local fluctuations of the magnetic order parameter. 
Our many-body calculations on an \emph{ab initio}-derived multi-orbital basis predict indeed a kinetic energy gain when entering the magnetically ordered phase.
La$_2$NiTiO$_6$ emerges thus as a paradigmatic realization of a Hund's coupling-driven Mott insulator. Its peculiar properties may turn out to be instrumental in the ongoing chase after correlated topological states of matter.  
\end{abstract}


\maketitle

\let\n=\nu \let\o =\omega \let\s=\sigma

\vskip 5mm

\noindent
\section{Introduction} 
Nickel (Ni) in $d^8$ configuration has been attracting growing attention for the possibility to realize the ``Haldane'' $S$=1 spin-chain \cite{haldanePRL50,affleckPRL59,buyersPRL56,whitePRB48,golinelliPRB50,cizmarNJP10}. In compounds like CsNiCl$_3$ or NiTa$_2$O$_6$ the Ni atoms are connected via small hopping integrals $t$ along specific one-dimensional paths and charge fluctuations are strongly suppressed by the large on-site Hubbard repulsion $U$. This allows for a theoretical description in terms of the 1D-Heisenberg model with an antiferromagnetic superexchange coupling $J \propto t^2/U$. 
In two dimensions the interest in $S$=1 quantum antiferromagnets has been somewhat hidden by the widely investigated spin-$1/2$ $t$-$J$ model, related to the physics of underdoped high-$T_c$ cuprates. Ni is again present in some of the $S$=1 bulk materials with strong 2D character, such as La$_2$NiO$_4$ or K$_2$NiF$_4$ \cite{birgeneauPRB16,grevenZPhysB96,khuntiaJPhysCM22,tsujimotoChemMat22,birgeneauPRB41}.
In an interesting recent proposal Chen, {\it et al.} suggested to artificially design a 2D spin-1 Mott insulator upon heterostructuring Ni and Ti single perovskites \cite{chenPRL111}.

In 3D spin-1 quantum magnets are found in pyrochlore compounds, such as ZnV$_2$O$_4$ or MgV$_2$O$_4$ \cite{yamashitaPRL85,garcia-adevaPRL85,gardnerRMP82}, where the absence of magnetic ordering down to very low temperatures is, however, due to frustration rather than to the strong-coupling regime in $U$.  
Some face-centered cubic (fcc) transition-metal oxides with $S$=1 are described in terms of spin-only models with nearest- (90$^\circ$) and next-nearest neighbor (180$^\circ$) exchange couplings $J_1$ and $J_2$, respectively. While this is fully justified for NiS$_2$ \cite{matsuuraPRB68,perucchiPRB80}, which belongs to the family of frustrated magnets ($J_2/J_1 \approx 0.5$), NiO \cite{kunesPRL99} and KNiF$_3$ \cite{martinPRL79}, together with $d^2$-vanadates \cite{deRayPRL99,note-d3}, are actually quite far from the strong-coupling Heisenberg limit, due to the significant hybridization between the transition-metal ions and the ``bridging'' ligand atoms.
Charge fluctuations indeed still play a role as also reflected by the relevant $d$-electron bandwidth, which in these compounds hardly gets smaller than $\sim$1.5-2.0\,eV.
As a matter of fact, the majority of the spin-1 three-dimensional transition-metal compounds that we know of, fall into one or both of the following categories: materials with relatively high magnetic ordering temperatures and pretty far from a true strong-coupling Heisenberg limit, or quantum magnets where long-range order is suppressed by sizable geometrical frustration.
The examples that are lacking for $S$=1 in 3D are those of nearly unfrustrated cases with small values of the ratio $t/U$, i.e. the repulsive counterpart of phase-fluctuation driven Bose-Einstein physics. In such materials, the strong coupling regime would determine the low magnetic ordering temperatures $T_\text{N} \propto J$. 

Here we demonstrate that the Nickel double perovskite La$_2$NiTiO$_6$ is a perfect realization of the latter class of systems. 
As we show in our calculation, the reason why this $S$=$1$ quantum antiferromagnet is deep into the Heisenberg limit comes from its distinctive hierarchy of magnetic exchange couplings: $J_2 \gg J_1$. La$_2$NiTiO$_6$ can therefore be very well described by $S$=$1$ spins living on a weakly frustrated three-dimensional fcc lattice \cite{linesPR139th,yildirimPRB58,ignatenkoJETP87}.
In order to fully describe the residual charge fluctuations, which in spin-1 systems may be relevant due to the importance of biquadratic effects as well as three-body interactions \cite{sutherlandPRB12,yoshimoriJPSJ50,aletPRB83,michaudPRB88}, we also go beyond the bilinear spin-only description and investigate the antiferromagnetic (AFM) phase in the ``full'' Hubbard model. This allows us to make a thermodynamic analysis of La$_2$NiTiO$_6$ revealing a kinetic-energy driven ordering mechanism.

The low value of the N\'eel temperature $T_\text{N}\sim 25$K \cite{rodriguezJMC12} in La$_2$NiTiO$_6$ has the interesting consequence that its paramagnetic Mott insulating state can be observed in an unusually extended range of temperatures.
Even though its properties as a Mott insulator have not been discussed hitherto, it is important to stress that La$_2$NiTiO$_6$ can actually be synthesized, as described in Refs. \onlinecite{royJAmCeramSoc37,fratelloJCG166,rodriguezJMC12,floresJMC21,yangJAP111}.  
Here we connect its features as a high-spin paramagnet with the peculiar electronic structure: a half-filled $e_g$ manifold at the Fermi level which is extremely narrow and uncommonly well separated from any other band.
The origin of this lies in the isotropic reduction of the hoppings in all three spatial directions, something hardly possible to achieve artificially but that nature does very effectively, replacing the Ni-O-Ni bonds characteristic of other $S$=$1$ materials with longer Ni-O-Ti-O-Ni ones.
This class of $d^8$-$d^0$ double perovskites can open new directions in oxide engineering: by considering also heavier elements of the Ni group and upon splitting the $e_g$ bands by heterostructuring or strain a correlation-driven band inversion can be realized, as in recent theoretical proposals for interacting topological insulators \cite{raghuPRL100,pesinNaturePhys6,fuPRL106,rueeggPRB88,budichPRB87}.

\begin{figure}[t]
  \includegraphics[width=0.8\columnwidth]{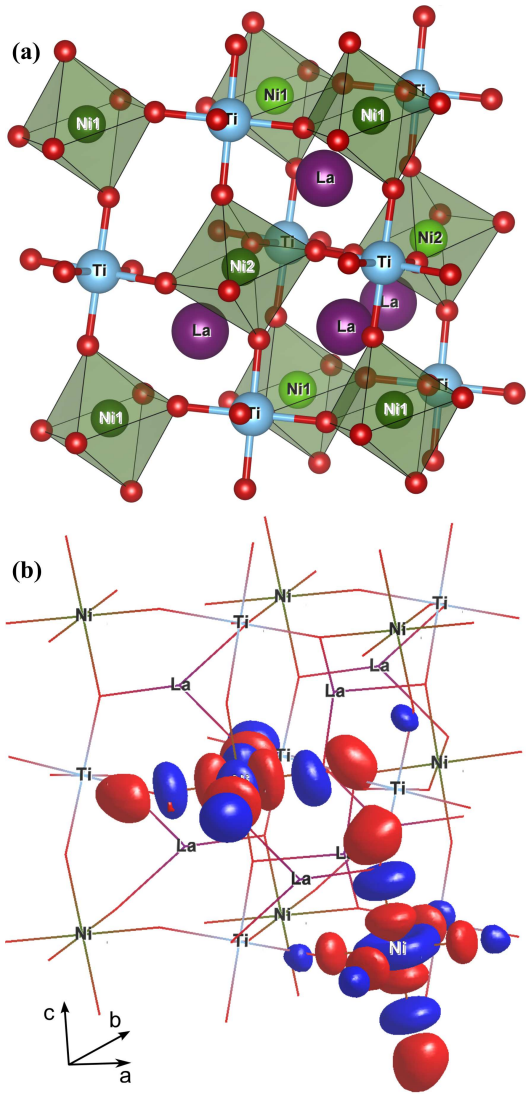}
  \caption{(color online) (a) Crystal structure of La$_2$NiTiO$_6$. (b) Isosurfaces of Wannier functions obtained by projection of only the Ni $e_g$ bands. The upper left one is mainly of $x^2-y^2$ character and the lower right mainly of $3z^2-r^2$ character. The coordinate system gives the directions used in Tab.~\ref{tab:hopping} and also applies to panel (a).}
  \label{fig:struct}
\end{figure}

\section{Electronic structure}
La$_2$NiTiO$_6$ crystallizes in a double perovskite structure with a small monoclinic distortion ($P2_1/n$ space group), as determined from neutron powder diffraction experiments \cite{rodriguezJMC12,floresJMC21,yangJAP111}. Structural relaxation within Density Functional Theory (DFT) using the GGA (PBE) \cite{PBE} functional results only in minor changes to the experimentally measured structure. The Ni-Ni distances along the $\bm{a}$ and $\bm{b}$ axes are $7.85$\,\AA, while along $\bm{c}$ the distance is $7.83$\,\AA.
The Ni/TiO$_6$ octahedra display a very small Jahn-Teller distortion (the lengths of the Ni/Ti-O bonds differ by at most 0.4\%) and show an alternating tilting (see Fig. \ref{fig:struct}a). 

For the paramagnetic calculations we consider a unit cell containing two formula units whereas the magnetic cell contains four. The locally equivalent Ni atoms form an fcc lattice comprised of intertwined simple tetragonal sublattices (denoted as ``1'' and ``2'' in Fig. \ref{fig:struct}a). One ``face'' of the fcc lattice formed by the darker green (darker grey) Ni atoms is shown in Fig. \ref{fig:struct}a. A face with Ni1 corners has a Ni2 in the center and vice versa. 

\begin{figure}[t]
  \includegraphics[width=\columnwidth]{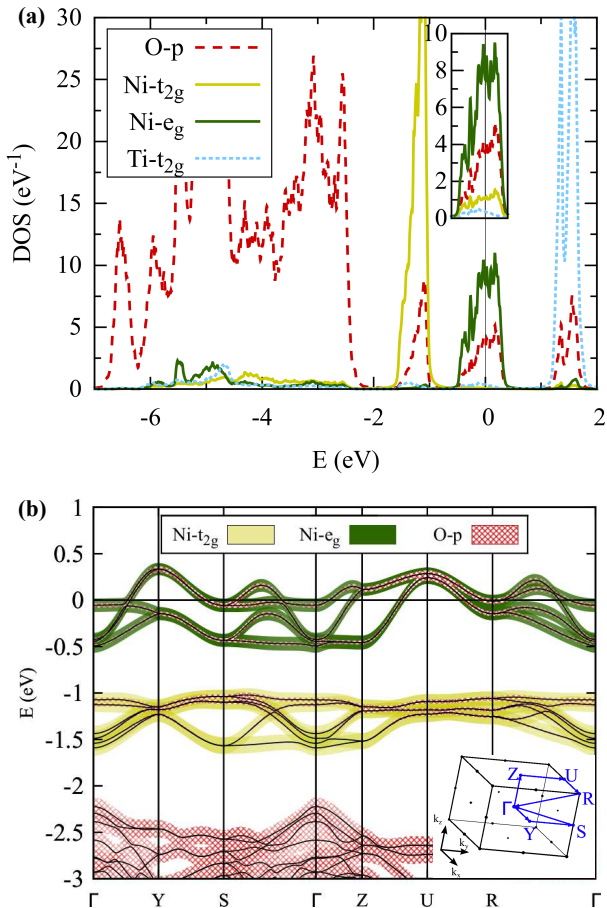}
 \caption{(color online) (a) Orbitally resolved density of states (Fermi level at $E\!=\!0$) (b) Fat band electronic structure for a cell containing two formula units. The thickness of the bands denotes the corresponding orbital character. In both panels the Wannier functions that have been used are those that diagonalize the occupancy matrix spanning Ni $d$, Ti $t_{2g}$ and O $p$ states (see text). In this basis the character mixture between the $t_{2g}$ and $e_g$ states of Ni is almost absent, in contrast to the MLWF basis, an example of which is shown, for the DOS, in the inset to panel (a).}
  \label{fig:DOS}
\end{figure}

The electronic structure was calculated with GGA using the \textsc{vasp} code \cite{VASP}. Nominally Ni is in a $3d^8$ configuration and Ti in $3d^0$. In DFT La$_2$NiTiO$_6$ is a metal, with two degenerate Ni $e_g$ bands crossing the Fermi level, as shown in Figs. \ref{fig:DOS}a and b. 
Due to the presence of the inactive Ti ``spacers'' the Ni $e_g$ bands are remarkably narrow. The corresponding value of the bandwidth $W_{e_g} \sim 0.8$\,eV is indeed substantially smaller than that of NiS$_2$ \cite{schusterEPJB85,perucchiPRB80,kunesPRB81}, of NiO \cite{mattheissPRB5} and of other $S$=$1$ three-dimensional compounds.
The $t_{2g}$ manifold of Ni lies 1\,eV below the Fermi level and, approximately 1\,eV further below, one finds the upper edge of the O $2p$ bands. 
The states close to the Fermi level are predominantly of Ni $e_g$ character and are furthermore well separated from the other bands.

Subsequently we extracted maximally-localized Wannier functions (MLWF) \cite{wannierRMP} from the O $2p$, the Ni $3d$ as well as the Ti $t_{2g}$ bands using the \textsc{wannier90} package \cite{wannier90}. Due to the tilting and rotation of the octahedra the straightforward MLWF construction produces a basis that retains considerable on-site mixing between the Ni $t_{2g}$ and $e_g$ orbitals (see inset of Fig. \ref{fig:DOS}a). This local $t_{2g}$-$e_g$ hybridization is just a consequence of this specific choice of orbital representation, therefore we have performed a unitary transformation after the MLWF procedure \cite{lechermann2006}. The usual choices here are a rotation into the so-called ``crystal field basis'' or into a basis that renders the DFT occupancy matrix $\rho_{ij}=\langle c^\dagger_{i} c^{\phantom{\dagger}}_j\rangle$ 
diagonal on each atom, see e.g. Refs. \onlinecite{lechermann2006,pavariniNJP7}.
In light of a subsequent dynamical mean field theory (DMFT) calculations using a quantum Monte Carlo solver we have decided to block diagonalize the occupancy matrix, since this treatment yields in our case smaller off-diagonal elements in the frequency-dependent non-interacting Green's function $G^0(i\omega_n)$ as the crystal field basis. We quantify the off-diagonal elements in $G^0(i\omega_n)$ by the average of the absolute values of the off-diagonal elements, i.e.
\begin{equation}
  \overline{\left\vert G^0(i\omega_n) \right\vert} = \frac{1}{N_{\rm od}} \sum_{m>m'}\left\vert G^0_{mm'}(i\omega_n) \right\vert,
\end{equation}
where $N_{\rm od}$ is the number of off-diagonal elements in the upper triangle of the matrix. We find the largest values at the first Matsubara frequency, for the crystal field basis and inverse temperature of $\beta=40\mathrm{eV}^{-1}$ the value is $\overline{\left\vert G^0(i\omega_0) \right\vert}=0.06\mathrm{eV}^{-1}$, while for the diagonal occupancy matrix the same is more than a factor of five smaller at $0.01\mathrm{eV}^{-1}$.
The orbital character shown in the density of states in Fig. \ref{fig:DOS}a and the band structure in Fig. \ref{fig:DOS}b was computed using this basis. 

Because of the separation of the states close to the Fermi level from the other bands and their predominant Ni $e_g$ character we construct a low-energy model using only these bands, projecting onto MLWFs spanning this subspace. This results in two orbitals sitting on Ni that are warped from the atomic shape by hybridization with O and Ti, an $x^2-y^2$-like and a $3z^2-r^2$-like Wannier function, whose isosurfaces are shown in Fig.~\ref{fig:struct}b. In this case no additional basis transformation was necessary, since the MLWFs are already locally orthogonal. This two band model is used for most of the DFT+DMFT calculations presented here. A larger basis containing the full Ni $d$ and the O $p$ shell was also considered within DFT+DMFT for assessing the validity of the two band description, see Section \ref{sec4}.

The calculated Ni-Ni hopping amplitudes in this $e_g$-only model for the $3z^2-r^2$- and $x^2-y^2$-like orbitals, effectively containing the hybridization to O and Ti, are shown in Tab. \ref{tab:hopping}. We label the orbitals as $\ket{1}\sim 3z^2-r^2$ and $\ket{2}\sim x^2-y^2$ on Ni1 and analogously $\ket{3},\ket{4}$ on Ni2. The hopping amplitude between orbitals $\ket{i}$ and $\ket{j}$ in a given direction is given by $t_{i,j}$. The overall Ni-Ni hopping is small, the element $t_{1,1}$ along the $\bm{c}$ axis being the largest  ($-97$\,meV). Along the same direction the $x^2-y^2$ hardly contributes. 
In the $\bm{ab}$-plane the situation is more evenly distributed among the two orbitals but the sum of the squares of all hoppings is similar to the same quantity along $\bm{c}$ (as shown in Eq. \ref{Heff} $\sum_{mm'}|t_{mm'}|^2$ determines the super-exchange coupling). 
For Ni-Ni 90$^\circ$ bonds there are two possible paths, either inter- or intra-sublattice hoppings, i.e. either Ni1-Ni2 or Ni1-Ni1, respectively.
The most important outcome of the Wannier projection is that the next-nearest-neighbor Ni-Ni 180$^\circ$ hoppings are a factor of 4 to 10 larger than the nearest-neighbor 90$^\circ$ ones.


\section{Spin-only model}
\begin{figure}[t]
  \includegraphics[width=0.95\columnwidth]{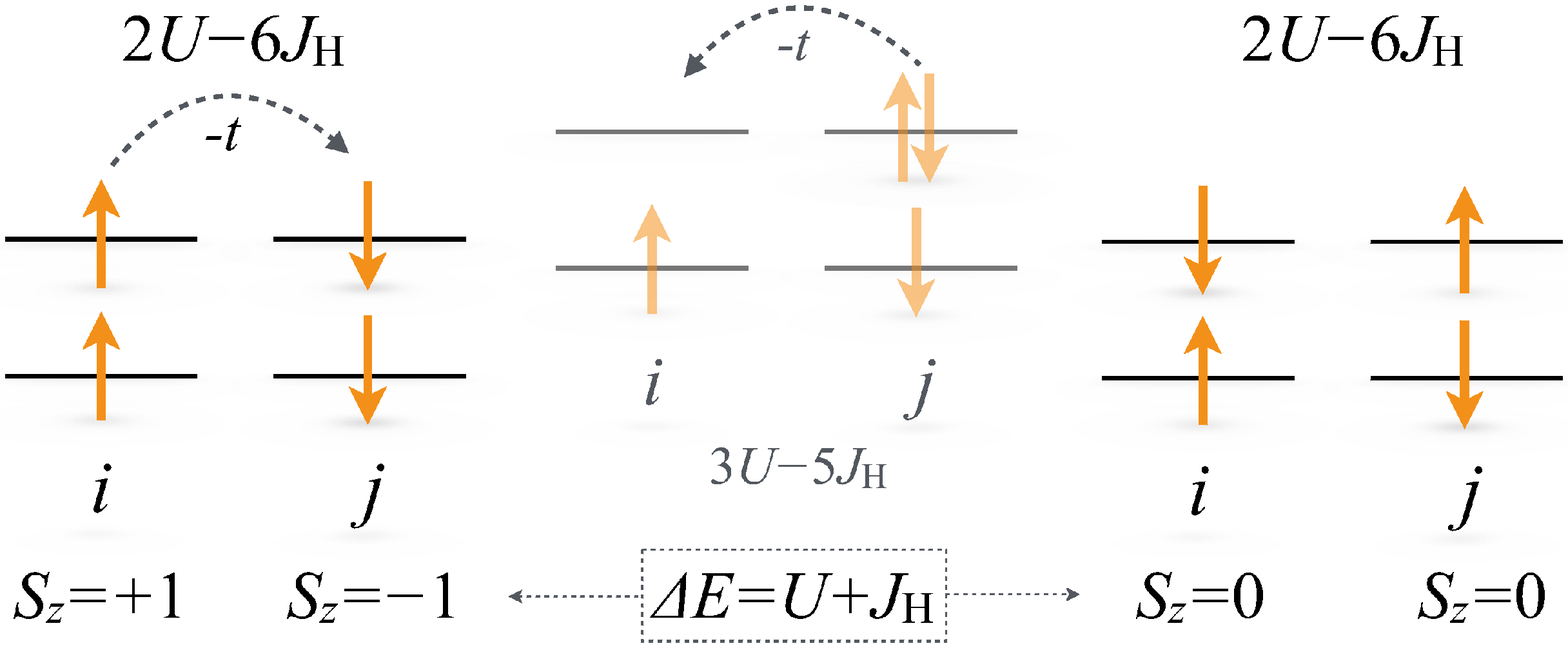}
  \caption{(color online) Sketch of a spin-spin off-diagonal term of $\mathcal{H}_{\rm{Heis.}}$ between neighboring sites $i$ and $j$. Even though the two $e_g$ orbitals are shown on two different levels for the sake of clarity, we stress that they are in fact degenerate. The final state is actually the triplet combination which, for simplicity, is represented as just one state.}
  \label{fig:scheme}
\end{figure}

The Wannier projection allows us to derive a bilinear Heisenberg Hamiltonian $\mathcal{H}_{\rm {Heis.}}$, with which we can give a first description of the physics of La$_2$NiTiO$_6$. 
To this aim, we downfold \cite{loewdin51,pavariniNJP7} our \emph{ab initio} $e_g$ model with two electrons onto the subspace of singly occupied orbitals. The intermediate configurations generated by one Ni-Ni hopping process contain only one intra-orbital double occupation, as sketched in Fig. \ref{fig:scheme}.
The local interaction considered in the direct exchange model is of Kanamori type~\cite{kanamori,georgesCMP137}, the fully SU(2)-symmetric interaction Hamiltonian reads

\begin{equation}
\hspace*{-7pt}
\begin{aligned}
\mathcal{H}_{\rm Kan.}=&\phantom{+}U \sum_{m} n_{m,\uparrow} n_{m,\downarrow} \\
&+\sum_{\substack{m>m^\prime\\\sigma}} \left[U^\prime n_{m,\sigma} n_{m^\prime,-\sigma}+(U^{\prime}-J_{\rm H}) n_{m,\sigma} n_{m^\prime,\sigma}\right] \\
&+\frac{1}{2}J_\text{H}\sum\limits_{\substack{m \neq m^\prime\\ \sigma}} \Bigl(c^\dagger_{m,\sigma}c^\dagger_{m^\prime,-\sigma}c^{\phantom{\dagger}}_{m,-\sigma}c^{\phantom{\dagger}}_{m^\prime,\sigma} \\
&\hspace{70pt}-c^\dagger_{m,\sigma}c^\dagger_{m,-\sigma}c^{\phantom{\dagger}}_{m^\prime,\sigma}c^{\phantom{\dagger}}_{m^\prime,-\sigma}\Bigr)
\end{aligned}
\label{eq:hamiltonian}
\end{equation}
\vspace*{2ex}

\noindent with the number operator $n_{m,\sigma}=c^\dagger_{m,\sigma}c^{\phantom{\dagger}}_{m,\sigma}$, where $c^\dagger_{m,\sigma}$ ($c^{\phantom{\dagger}}_{m,\sigma}$) creates (annihilates) an electron with spin $\sigma$ in orbital $m$.
Furthermore, we used $U^\prime=U-2J_\text{H}$, where $U$ represents the Hubbard repulsion and $J_\text{H}$ the Hund's coupling.

The two electrons on each Ni give rise to $S$=1 as well as $S$=0 configurations which, in the low-energy subspace, are mutually coupled. However, as we will see later in our dynamical mean field theory calculation, the two electrons are strongly affected by the Hund's coupling $J_{\text{H}}$ and yield an effective local moment close to the maximum possible value. 
Hence, the singlet can be discarded from our analysis. The resulting $S$=1 Heisenberg Hamiltonian reads
\begin{equation} \label{Heff}
\mathcal{H}_{\rm {Heis.}} = \frac{1}{U+J_{\rm H}} \big(\sum_{mm'}|t_{mm'}|^2 \big) \sum_{ij} \big( \bm{S}_i \cdot \bm{S}_j - 1 \big).
\end{equation}
One of the processes responsible for the spin off-diagonal terms is shown in Fig. \ref{fig:scheme}, where also the energies of the initial/final and intermediate states are given. The initial and intermediate configurations are eigenstates of the Kanamori Hamiltonian. The final state is actually the triplet combination which, for simplicity, is represented as just one state in our sketch.

\begingroup
\squeezetable
\begin{table}
\begin{ruledtabular}
\begin{tabular}{cccccc}
$abc$ & \multicolumn{4}{c}{hopping amplitude $t_{ij}$ (meV)} & $\sum\limits_{i,j}t_{ij}^2$ (meV$^2$)\\
\hline
 & (1 1) & (1 2) & (2 1) & (2 2) & \\
\hline
0 0 1 & -97 & -3 & -3 & 0 & 9427\\
0 1 0 & -27 & 44 & 44 & -70 & 9501\\
1 0 0 & -22 & -42 & -42 & -77 & 9941\\
\hline
 & (3 3) & (3 4) & (4 3) & (4 4) & \\
\hline
0 0 1 & -97 & -3 & -3 & 0 & 9427\\
0 1 0 & -22 & -42 & -42 & -77 & 9941\\
1 0 0 & -27 & 44 & 44 & -70 & 9501\\
\hline
 & (1 3) & (1 4) & (2 3) & (2 4) & \\
\hline
0 1 1 & -27 & 25 & -19 & -4 & 1731\\
1 0 1 & -27 & 25 & -19 & -4 & 1731 \\
0 $\bar{1}$ 1 & -24 & 26 & -18 & 0 & 1576\\
1 0 $\bar{1}$ & -24 & 26 & -18 & 0 & 1576\\
\hline
 & (1 1)/(3 3) & (1 2)/(3 4) & (2 1)/(4 3) & (2 2)/(4 4) & \\
\hline
1 1 0 & 11 & 2 & 2 & -46 & 2245\\
1 $\bar{1}$ 0 & 12 & 3 & 3 & -26 & 838\\
\end{tabular}
\end{ruledtabular}
\caption{Hopping parameters between two Ni atoms within the crystal as obtained via Wannier projection. The first column indicates the direction of the Ni-Ni bond via $\bm{v}\propto a\bm{a}+b\bm{b}+c\bm{c}$. The numbers in parentheses refer to the indices $i, j$ by $\ket{1}\sim 3z^2-r^2$ and $\ket{2}\sim x^2-y^2$ on Ni1 and analogously $\ket{3},\ket{4}$ on Ni2. Only hopping amplitudes between nearest and next-nearest Ni atoms are given here.}
\label{tab:hopping}
\end{table}
\endgroup

\begin{figure}[t]
  \includegraphics[width=\columnwidth]{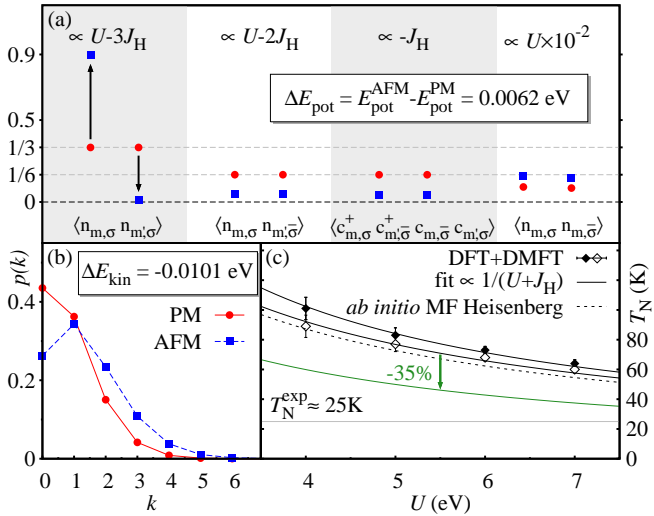}
  \caption{(color online) Energetic balance for $U\!=\!4$\,eV, $J_{\rm H}\!=\!0.6$\,eV and $\beta\!=\!200$\,(eV$)^{-1}$. (a) Different contributions to the potential energy of the paramagnetic and antiferromagnetic phases (red dots and blue squares, respectively). The two sets of data given per quantity correspond to the two spin orientations. The error bars are not visible as they are smaller than the symbol size. The potential energy of the AFM solution is larger than that of the PM one (potential energy loss). (b) Histogram of the expansion order of the QMC diagrams contributing to the fermionic trace for the two phases. Its average is proportional to the kinetic energy. The shift towards higher expansion orders for the AFM solution indicates a kinetic energy gain. In (c) $T_N$ calculated within DFT+DMFT for the $e_g$-only model is reported with diamonds (full and empty symbols correspond to $J_{\rm H}\!=\!0.6$ and $1.0$\,eV, respectively). Fits to the data (black solid lines) yield a prefactor of the $1/(U+J_{\rm H})$ behavior which is very close to the estimate obtained from the 
mean-field solution of a Heisenberg model with the hopping values from Table \ref{tab:hopping} (black dashed line, for $J_{\rm H}\!=\!1.0$\,eV). From Ref. \onlinecite{linesPR139th} we estimated the reduction of the mean-field value, due to spatial fluctuations (line indicated by the arrow).} 
  \label{fig:energy}
\end{figure}

Using typical interaction values for Ni ($U=5$\,eV and $J_{\rm H}=0.8$\,eV \cite{sasiogluPRB83}) we get $J_2 \simeq 1.6$\,meV and $J_1 \simeq 0.3$\,meV (or smaller, depending on which 90$^\circ$ bond is considered). This value of $U$ is moderate, since, for example, in NiO $U=8$eV \cite{anisimov91}.
This small value of the ratio $J_1/J_2 \sim 0.2$ -- a direct consequence of the small nearest-neighbor hoppings -- corresponds to a very weak degree of frustration. The 180$^\circ$ Ni-Ni bonds are not strongly disturbed by the nearest-neighbor ones and form four interpenetrating antiferromagnetic simple cubic sublattices. 
The magnetic ordering vector of this so-called AF-II phase, which in mean-field is stable for $J_1 < 2 J_2$, is [$\nicefrac{1}{2}$,$\nicefrac{1}{2}$,$\nicefrac{1}{2}$] \cite{linesPR139th,yildirimPRB58,ignatenkoJETP87}.
We have performed GGA+U calculations and found that the AF-II (Type A) order has indeed the lowest energy, in agreement with experiments \cite{rodriguezJMC12,martin_master}.

\section{DFT+DMFT calculation}\label{sec4}
In order to go beyond the spin-only bilinear Heisenberg model above, we solve the ``full'' multi-orbital Hubbard model in the Wannier basis using dynamical mean field theory \cite{dmft_1,dmft_2,dmft_3}.
In the following we present calculations for the $e_g$-only basis with the SU(2)-symmetric Kanamori interaction. 
The result is that La$_2$NiTiO$_6$ is a Mott insulator in DFT+DMFT. 
We have also tried larger basis-sets, in particular a $dp$-model containing Ni $e_g$, Ni $t_{2g}$ and O $p$ bands. 
The DFT+DMFT result turns out to be robust against the choice of the low-energy model, in contrast to many other transition-metal oxides for which DFT+DMFT gives qualitatively different outcomes depending on the basis set \cite{dp-papers}.
In selected cases we performed calculations for an enlarged model containing the full Ni $d$ and the O $p$ shells, using density-density (only the first two lines of Eq. \ref{eq:hamiltonian}) as well as Kanamori interaction. As a result we find that the system is still a Mott insulator with a Ni $d$ occupation of about 8.5 electrons, i.e. 2.4 electrons in the $e_g$ states.
We note in passing that this robustness of La$_2$NiTiO$_6$ against the choice of basis set makes it an ideal testbed material for the derivation of low-energy models for $e_g$ orbitals, in the same way as SrVO$_3$ is very often used for $t_{2g}$ bands. La$_2$NiTiO$_6$ has the additional interesting property of a much stronger effect of the Hund coupling $J_\text{H}$ because of the half-filled, narrow $e_{g}$ bands.


The DFT+DMFT solution of La$_2$NiTiO$_6$ for the $e_g$-only model demonstrates that, in a wide range of interaction parameters relevant for Ni ($U=4\,$eV to $7$\,eV and $J_{\rm H}=0.6-1.0$\,eV), the local moment is very close to the maximum value of $S_{\text{eff}}=1$. By calculating $\langle S_z^2 \rangle$ we indeed find its maximum value of 2/3, because the inter-orbital ``Hund'' double occupancies $d_\text{H}$=$\langle n_{1,\uparrow} n_{2,\uparrow} \rangle$ and the ``anti-Hund'' ones $d_{\text{anti-H}}$=$\langle n_{1,\uparrow} n_{2,\downarrow} \rangle$ are given by their ``saturation'' values of $1/3$ and $1/6$, respectively (see Fig. \ref{fig:energy}a). In the paramagnetic phase we therefore have 
$\langle {\bm{S}}^2 \rangle = 3 \langle S_z^2 \rangle = 2 = S_{\text{eff}} (S_{\text{eff}}+1)$ 
with the SU(2)-symmetric Kanamori interaction. Hence $S_{\text{eff}}=1$ and, as we only consider its spin-dependent contribution ($g=2$), the corresponding local moment is $m \simeq 2.83\mu_{\rm B}$.

So far we have used DFT+DMFT to analyze the paramagnetic phase of La$_2$NiTiO$_6$. 
Being a mean-field theory, DMFT allows us to follow it down to zero temperature or, alternatively, to calculate the N\'eel temperature and switch to the magnetically ordered solution below $T_{\rm N}$.
The values of $T_{\rm N}$ calculated in our \emph{ab initio} $e_g$-only model for different values of $U=4$\,eV to $7\,$eV are shown by the full and empty diamonds in Fig. \ref{fig:energy}c for $J_\text{H}\!=\!0.6$\,eV and $1.0$\,eV, respectively. 

Before making a close comparison between the DMFT results and the experimental $T_{\rm N}$ some considerations are in order:
Even if DFT+DMFT is well known for giving accurate results for three-dimensional transition-metal oxides, the quantitative corrections due to spatial fluctuations are still sizeable in 3D. 
The reduction of $T_{\rm N}$ is one of the most evident of these corrections. Indeed, even if not as dramatic as in 2D, where the single-site DMFT $T_{\rm N}$ is finite instead of zero as predicted by the Mermin-Wagner theorem, this reduction has been quantified by means of a diagrammatic extensions of DMFT to be $\sim$30\% in the intermediate-to-strong coupling regime \cite{rohringerPRL107}. 

In our specific case, we can also rely on random-phase calculations and on spin-wave theory to evaluate the effect of non-local correlations. For our value of the $J_1/J_2$ ratio the random-phase approximation predicts for the fcc case with $S$=$1$ a reduction of $T_{\rm N}$ of about 35\% compared to mean-field \cite{linesPR139th}, in line with the above-mentioned result.
The solid line indicated by the arrow in Fig. \ref{fig:energy}c represents the DFT+DMFT results taking into account the 35\% reduction. This line gets quite close to the experimental value, especially for the largest values of $U$ considered.
The most plausible reason for an additional reduction of the theoretical $T_{\rm N}$ is the presence of a few percents of Ni-Ti anti-site disorder, as reported in Ref. \onlinecite{rodriguezJMC12,floresJMC21,yangJAP111}.

Before switching to the thermodynamics of the magnetic transition, let us also comment on the dashed line in Fig. \ref{fig:energy}c. This shows the behavior with $U$ of the mean-field N\'eel temperature of a $S$=$1$ Heisenberg model on an fcc lattice ($k_B T_{\rm N} = 4 J_2$, see Ref. \onlinecite{linesPR139th}), where in the expression for $J_2$ the hopping values estimated from our DFT analysis have been used (as in Eq. \ref{Heff}). The almost perfect agreement with $T_{\rm N}$ from the full DFT+DMFT calculation shows that it makes perfect sense to identify the single-site DMFT result with the mean-field Heisenberg outcome. 

In order to prove that the physics of La$_2$NiTiO$_6$ is actually that of a strong-coupling Heisenberg antiferromagnet, we perform a thermodynamic analysis.
The smoking gun ruling out possible intermediate-coupling physics is a lower total energy for the AFM phase realized through a kinetic energy gain and a loss in potential energy \cite{toschiPRB72,tarantoPRB85,gullEPL84,schaefer2014}. 
Our results very clearly indicate a kinetic energy gain, as shown in Fig. \ref{fig:energy}b. This is calculated from the first moment of $h(k)$, the histogram of the expansion order of the continuous-time hybridization-expansion quantum Monte Carlo solver \cite{CTQMC,noteCTQMC}. A shift toward larger expansion orders indicates a gain in kinetic energy for the AFM phase ($\Delta E_{\text{kin}}=E^{\text{AFM}}_{\text{kin}}-E^{\text{PM}}_{\text{kin}}<0$).
At the same time, as shown in Fig. \ref{fig:energy}a, where the local terms of the multi-orbital Hubbard Hamiltonian are separately analyzed, we detect a potential energy loss ($\Delta E_{\text{pot}}=E^{\text{AFM}}_{\text{pot}}-E^{\text{PM}}_{\text{pot}}>0$), ruling out intermediate-coupling physics. 
Our analysis reveals that the potential energy loss is almost entirely given by the corresponding increase in $d_U$, the intra-orbital double occupancies by going from PM to AFM. 
As shown in Fig. \ref{fig:energy}a, the ``Hund'' inter-orbital double occupancies $d_\text{H}$ (proportional to $U\!-\!3J_{\rm H}$), the ``anti-Hund'' ones, $d_{\text{anti-H}}$ ($\propto U\!-\!2J_{\rm H}$) and the ``spin-flip'' term $d_\aleph$ ($\propto -\!J_{\rm H}$), are close to compensating each other. Since the pair-hopping terms hardly contribute, the potential energy loss reads $\Delta E_{\text{pot}}\!=\!2[(U\!-\!3J_\text{H})\Delta d_\text{H} + (U\!-\!2J_\text{H})\Delta d_{\text{anti-H}} + J_\text{H}\Delta d_\aleph + U \Delta d_U]$, where $\Delta d$ indicates the total difference in the respective quantity summed over spin and orbital indices.
Indeed it is almost entirely given by the corresponding change in $d_U$: $\Delta E_{\text{pot}} \approx 2U\Delta d_U$.

This is a precise consequence of the strong-coupling physics: the disordered phase has preformed localized moments that slightly delocalize upon entering the ordered phase because they gain coherence. At the same time, the length of the (unordered) local moment (whose square is $\propto \langle {\bm{S}}^2 \rangle$ \cite{note_DeltaEpot}) changes only slightly by going from the PM to the AFM phase: the latter is $\sim 0.002\mu_{\rm B}$ shorter than the former.

\section{Conclusions}
We have shown that La$_2$NiTiO$_6$ is a Hund's coupling-driven Mott insulator, far in the strong-coupling limit. 
The peculiar properties of this double perovskite come from the presence of inactive $d^0$ Ti-``spacers'' which enlarge the Ni-Ni bonds isotropically in all directions, drastically reducing the relevant bandwidth.
The proper low-energy spin-spin model is a Heisenberg Hamiltonian with next-nearest-neighbor exchange coupling $J_2$ equal to about 1.6\,meV and nearest-neighbor coupling $J_1$ a factor of 4-5 smaller. On an fcc lattice like the present one, this implies that frustration effects are almost absent and the very low value of $T_{\rm N}$ is a consequence of the strong local fluctuations of the order parameter. 
We demonstrate the strong coupling nature of La$_2$NiTiO$_6$ by a direct analysis of the energetic balance within DFT+DMFT. This prediction can be tested, for instance, by looking for the presence of spin-polarons in photoemission as well as in optical conductivity measurements \cite{strackPRB46,sangiovanniPRB73,tarantoPRB85} which should be visible due to the pronounced three-dimensional character.

Our results unveil a new family of double perovskites -- La$_2$NiTiO$_6$ being its first member -- that, due to the dramatic reduction of the bandwidth can be very interesting for oxide engineering.
One promising direction is to try to split the two $e_g$ bands with strain or upon heterostructuring. This can be achieved because, despite the very isotropic $J_2$, the 180$^\circ$ hoppings of the $3z^2-r^2$- and of the $x^2-y^2$-orbitals are not symmetric under rotations of the crystal axes. 
It should therefore be possible to induce a splitting which, due to the hybridization between the two $e_g$-orbitals, may result in a gap of inverted orbital character at specific points of the Brillouin zone. The resulting band structure can in fact be ideal for the realization of a correlated topological insulator, with two $d$ electrons in two entangled orbitals forming a large local moment (hence more easily detectable in an experiment). If the $x^2-y^2$/$3z^2-r^2$ splitting turns out to be externally tunable, this class of $d^8$-$d^0$ double perovskites could become tremendously attractive from this point of view.
By substituting Ni with heavier isoelectronic elements the spin-orbit coupling can also help in the opening of the hybridization gap necessary to realize a correlated topological insulator.

\begin{acknowledgments} We thank A. Toschi for enlightening suggestions, Y. Motome, A. Katanin, R. Claessen, M. Sing, J. Kune\v{s}, M. Capone, G. Giovannetti for useful comments and A. Hausoel for his great work on the ``w2dynamics'' code. 
This work has been funded by the Deutsche Forschungsgemeinschaft through the research units FOR 1162 (M.K.) and FOR 1346 (G.S.).
\end{acknowledgments}


\begin{references}
\bibitem{note-d3}
The same is true, for instance, for the $d^3$ perovskites SrMnO$_3$ and SrTcO$_3$ \cite{mravljePRL108}.
\bibitem{noteCTQMC}
For our calculations we have used the W\"urzburg-Vienna code package ``w2dynamics'' \cite{parraghPRB86}.
\end{references}
\end{document}